# Broadband dielectric microwave microscopy on μm length scales

Alexander Tselev[*)] and Steven M. Anlage[a)]
*Center for Superconductivity Research, Department of Physics, University of Maryland, College Park, Maryland 20742-4111*

Zhengkun Ma and John Melngailis
*Department of Electrical and Computer Engineering, University of Maryland, College Park, Maryland 20742-3285*

## Abstract

We demonstrate that a near-field microwave microscope based on a transmission line resonator allows imaging in a substantially wide range of frequencies, so that the microscope properties approach those of a spatially-resolved impedance analyzer. In the case of an electric probe, the broadband imaging can be used in a direct fashion to separate contributions from capacitive and resistive properties of a sample at length scales on the order of one micron. Using a microwave near-field microscope based on a transmission line resonator we imaged the local dielectric properties of a Focused Ion Beam (FIB) milled structure on a high-dielectric-constant $Ba_{0.6}Sr_{0.4}TiO_3$ (BSTO) thin film in the frequency range from 1.3 GHz to 17.4 GHz. The electrostatic approximation breaks down already at frequencies above ~10 GHz for the probe geometry used, and a full-wave analysis is necessary to obtain qualitative information from the images.

---

[a)] Author to whom correspondence should be addressed.

# I. INTRODUCTION

The rapid development of nanotechnology and the shrinking size of the functional elements of devices demand diverse tools for comprehensive materials characterization at the micron and sub-micron length scales. Near-field microwave microscopy has become a useful tool for imaging and measurement of electromagnetic properties of materials at the micron length scale.[1-6] Generally, the measurement principle of a microwave near-field microscope is based on alteration of the probe electrical impedance due to the presence of a sample in close proximity to the probe. To increase the microscope sensitivity, the probe is often made an integral part of a resonator – either distributed or based on lumped elements. Usually, due to fundamental limitations of a specific resonator design, the measurement frequency is limited to just one frequency or to a small frequency range. Little work has been done on broadband microwave microscopy. Chang *at al.*[7] performed measurements of permittivity on thin film $Ba_{1-x}Sr_xTiO_3$ (BSTO) phase spreads in a frequency range from 0.95 GHz to 4.95 GHz with the goal to investigate the compositional dependence of the BSTO permittivity dispersion. However, in addition to the frequency-dependent properties of materials, broadband measurements can also be implemented to separate and study the contributions of different sample properties in the measured microscope response. For instance, a broadband measurement can be used to disentangle dielectric permittivity and conductivity for a "leaky" dielectric, as well as to separate the influence of parasitic factors and systematic uncertainties accompanying, e. g., varying probe-sample coupling. Such an approach, called impedance spectroscopy, utilizing frequency sweeps is routinely used to study material properties at low frequencies with the use of LCR meters and impedance analyzers (see, e.g., Refs. [8,9]). One needs to cover at least one decade of frequency to obtain a reliable result of fitting to accomplish this task. Ideally, a microscope should acquire the full capability of an impedance analyzer while maintaining the ability to make spatially-resolved measurements.

In this respect, the microwave microscopes based on transmission line resonators have an advantage over microscopes based on cavity or lumped element resonators since the resonance mode structure in a transmission line resonator is maintained over a significantly broader frequency range, and the frequency sweep can be realized simply by jumping from one resonance mode to another. An electrically long transmission line resonator allows for a dense set of modes, creating a quasi-continuous frequency coverage.

Figure 1(a) shows the most general electrical schematic diagram of a transmission line resonator: the resonator – a segment of a transmission line of length $L$ – is coupled to a feed line with a coupling element of impedance $Z_c$ and terminated by a probe with effective impedance $Z_t$. Further, without significant loss of generality, we consider in more detail an example of a microscope with an electric probe based on a coaxial line resonator. Figure 1(b) displays, for this particular case, a schematic of the resonator open end with a small-radius apex tip acting as a probe in contact with a sample. Figure 1(c) shows a lumped element circuit representation of the probe tip and a lossy dielectric sample. Such a sample can be described by an $RC$-pair.[10,11] The capacitor, $C_s$, describes the fringe electric field energy stored in the sample, and the resistor, $R_s$, stands for all channels of energy dissipation in the sample. The sample is coupled to the resonator through the tip-sample coupling capacitance $C_{sc}$ and the resistance of the tip $R_{tip}$. The field between the tip and the outer conductor, which bypasses the sample, is represented by the parasitic stray capacitance $C_{str}$. The effective probe tip impedance $Z_t$ can be calculated as the impedance of the whole $RC$-network. The values of the equivalent circuit elements depend on the local properties of the sample under the probe tip and the set of parameters describing the tip-sample system. Generally, the dependence of the probe tip impedance on sample properties and the probe tip geometry is complicated and its elucidation constitutes the major problem of microwave microscopy.[2,6,12-16] A more detailed description of a sample will require a more complicated equivalent circuit.

When the tip is being scanned over the sample, the variation of the effective tip impedance causes shifts $\Delta f$ of the resonant frequencies and changes of the $Q$-factors of the resonator modes, which are monitored and imaged by a computer. Ideally, the broadband measurements could be accomplished by switching from one resonant frequency to another while the probe-sample geometry is fixed. By subsequent fitting of the microscope frequency-dependent response to a specific electrical model of the probe-sample system, the set of model parameters could be deduced, and further, the contributions of local properties of the sample could be separated and determined.

In this paper, we present an experimental effort of broadband imaging of a Focused Ion Beam (FIB) milled structure on a $Ba_{0.6}Sr_{0.4}TiO_3$ (BSTO) thin film, covering a frequency range from 1.3 GHz to 17.4 GHz with ~1 μm spatial resolution. Using a microscope system based on a coaxial transmission line resonator, we explore the possibility of distinguishing contributions of the film permittivity and the film conductance, particularly through the resonant frequency shift, by implementing broadband imaging.

## II. EXPERIMENT

A 370 nm-thick $Ba_{0.6}Sr_{0.4}TiO_3$ film was deposited by Pulsed Laser Deposition (PLD) at a substrate temperature of 800 °C at an oxygen pressure of 120 mTorr on a 500 μm-thick (001) $LaAlO_3$ (LAO) substrate. Milling of the BSTO film was performed with a focused 50 keV $Ga^+$ ion beam. The film was removed from a 5 μm-width trench around a 20 μm by 20 μm square island. As a result of the milling the material inside the trench as well as in some vicinity of the trench was doped with Ga. The structure was annealed at 600 °C in oxygen for 1 hour after milling to partially restore the damage caused by the milling. Such an annealing treatment has been shown to significantly improve the dielectric properties of the BSTO film in the damaged area.[17]

Basic details of the microscope system used in the present work were described elsewhere.[18,19] Here we summarize the additional features of the broadband microscope. The microscope is based on a half-wavelength coaxial transmission line resonator of length 25 cm with the inner conductor terminated by an Scanning Tunneling Microscope (STM) tungsten tip with about 1 μm-radius apex (Fig. 2). A microwave signal is generated by a synthesized microwave source (Agilent E8257C PSG Analog Signal Generator). The microwave power reflected from the resonator back into the feed line is monitored by a feedback circuit, which locks the frequency of the microwave source to the chosen resonance mode. At a resonance frequency, the reflection coefficient (Fig. 2(b)) passes a local minimum. The coupling of the feed line with the resonator can be accomplished by means of either a capacitor or an inductor coil inserted in the inner conductor between the resonator and the feed line. The best performance of the system can be achieved at modes close to the frequency of critical coupling of the resonator. (At the critical coupling, the reflection from the resonator at the resonant frequency is equal to zero). This condition dictated the use of several different coupling elements for broadband imaging, tuned for different frequency sub-bands. For the broadband experiments in this work, we used a modified SMA connector with miniature gold wire coils in place of the central conductor to couple the resonator with the feed line. Three coils with different inductances were used to cover the frequency range from 1 GHz to 17.5 GHz. Broadband imaging was performed by switching from one resonance mode of the microscope to another upon completion of the scanning frame at a previous frequency. The sample was maintained in constant contact with the probe tip by application of a small force of 30-50 μN to the sample.

For extraction of data on film permittivity, we used the approach outlined by Steinhauer *et al.*[19]. It is based on the perturbation formulas for resonant cavities and involves finite element calculations of fields around the probe tip in the air and in the sample in the electrostatic approximation. The procedure requires probe calibration against reference samples with known properties (e.g. bulk dielectric substrates). Refining that approach, in the present work, we more accurately described the geometry of the probe and used the ANSOFT Maxwell 2D finite element analysis package to evaluate the perturbation integrals. To reduce the size of the finite element model, the sample is placed on a conducting plane, which is considered to be at a floating potential in the numerical simulations.

## III. RESULTS AND DISCUSSION

*a) frequency shift images*

The upper panel of Fig. 3 shows a higher contrast and lower resolution frequency shift image over an 80 μm x 45 μm area of the FIB-milled structure, obtained by means of a blunter and therefore more sensitive tip. The image was taken at a frequency of 1.4 GHz. The FIB-milled trench looks like a broad black stripe (labeled "A" in Fig. 3) fringed with white lines ("B"). The bottom of the trench is heavily doped with Ga ions and relatively strongly conducting, producing a large frequency shift compared to the BSTO film. The small-frequency-shift white fringing is due to an "edge effect" when the tip apex is partially in contact with the film and partially hangs over the trench. The trench is surrounded on both sides by darker regions ("C") of Ga-doped BSTO film. The FIB beam has substantial tails that dope regions outside the main focus. The corners of the island surrounded by the trench ("D") are more heavily doped and are more conducting. Farther away is a broad brighter region of damaged film ("E") with a lowered dielectric permittivity producing a smaller frequency shift. A 1 μm-wide FIB-milled line ("F") is also seen along the left edge of the image. Bright spots ("G") result from holes in the film and particles of the PLD target material usually present on the surface of PLD-grown films.

The lower panel of Fig. 3 shows how the frequency shift contrast of the same region evolves with frequency. The image for $f = 2.2$ GHz looks similar to that for $f = 1.4$ GHz. At $f = 8.7$ GHz, the "edge effect" is still clearly seen, but the contrast between the trench bottom and the surrounding film is almost absent. There is no visible contrast between Ga-doped and undoped regions outside the trench. A slight contrast between the damaged area and the intact film is visible. At the highest frequency used in the experiments, $f = 17.4$ GHz, the trench looks white on a dark background. At this frequency, the frequency shift at the trench bottom is noticeably smaller than that for the surrounding and quite close to that corresponding to the LAO substrate without a film. A contrast between the damaged area and the intact film is still visible. It is worth noting here that images of film dielectric tunability (see Ref. [19] for the measurement method details) obtained simultaneously with the frequency shift images, all look similar to each other, regardless of frequency.

*b) theoretical model*

To interpret the transformation of the contrast of the images in Fig. 3, let us consider the relationship between the frequency shift and the sample properties in terms of the equivalent circuits in more detail. As mentioned above, the resonant frequency of the resonator is determined by measuring the microwave power reflected from the resonator back into the feed line. The complex reflection coefficient $\Gamma$ for voltage can be expressed through the reflection coefficients of the resonator coupling element $\Gamma_c$ and the terminating tip $\Gamma_t$ in the following form:[20]

$$\Gamma = \frac{\Gamma_c + \Gamma_t e^{-2\gamma L} - 2\Gamma_c \Gamma_t e^{-2\gamma L}}{1 - \Gamma_c \Gamma_t e^{-2\gamma L}}. \qquad (1)$$

Here, $\gamma = \alpha + j\frac{\omega}{c}\sqrt{\varepsilon}$ is the complex transmission line propagation constant, $L$ is the length of the resonator, $\omega$ is the angular frequency, $\omega = 2\pi f$, $c$ is the speed of the light in vacuum, $\alpha$ is the attenuation in the transmission line, and $\varepsilon$ is the relative permittivity of the transmission line dielectric. The expression Eq. (1) can be obtained by summing contributions from all partial waves multiply reflected in the resonator and contributing to the wave going back into the feed line.

The condition for the resonant minimum of the reflection coefficient:

$$\frac{d|\Gamma|^2}{d\omega} = 0, \quad \frac{d^2|\Gamma|^2}{d\omega^2} > 0 \qquad (2)$$

defines in implicit form the resonant frequency $\omega_r(c_i)$ as a function of a set of the parameters $c_i$ of the tip equivalent circuit. (For example, the set of $c_i$'s for the circuit in Fig. 1(c) is: $C_{sc}$, $C_s$, $C_{str}$, and $R_s$.) One is not interested in the value of $\omega_r$, but rather in the change of the resonant frequency $\Delta\omega_r$ caused by a change of the sample properties. Under the condition of a weak frequency dependence of the reflection coefficients $\Gamma_c$ and $\Gamma_t$:

$$f_0 \frac{\partial|\Gamma_{c,t}|}{\partial\omega} \ll 1, \quad f_0 \frac{\partial \mathrm{Arg}(\Gamma_{c,t})}{\partial\omega} \ll 1, \qquad (3)$$

it can be found by a straightforward calculation with use of Eqs. (2) that for a small perturbation caused by the sample:

$$\Delta\omega_r \approx \frac{\sum_i \partial\xi/\partial c_i \, dc_i}{\partial\xi/\partial\omega}, \qquad (4)$$

where $\xi = \mathrm{Arg}(\Gamma_c \Gamma_t e^{-2\gamma L})$, and $f_0 = \frac{c}{2L\sqrt{\varepsilon}}$ is the lowest mode frequency of a free half-wave resonator with length $L$. We note that $\xi$ is the change of the phase of a partial wave within the resonator for one round trip between the coupling element ant the tip. Equations (3) demand a weak frequency dependence of the reflection coefficients $\Gamma_c$ and $\Gamma_t$ within the resonant frequency shift for each resonant mode, and thereby set certain requirements to the design of the resonator-feed line coupling. Further, taking the derivatives of $\xi$ with account of the second of Eqs. (3), we find $\partial\xi/\partial\omega \approx 1/f_0$, $\sum(\partial\xi/\partial c_i)dc_i = \sum(\partial \mathrm{Arg}(\Gamma_t)/\partial c_i)dc_i$, and after that we obtain:

$$\Delta\omega_r \approx f_0 \sum_i \frac{\partial \mathrm{Arg}(\Gamma_t)}{\partial c_i} dc_i. \tag{5}$$

The complex reflection coefficient $\Gamma_t$ can be written in the form $\Gamma_t = \frac{1-z}{1+z}$ with

$z = Z_0 Y_t(\omega, c_i) = Z_0[G_t(\omega, c_i) + jB_t(\omega, c_i)]$. Here $Y_t = 1/Z_t$, $G_t$, and $B_t$ are the effective tip admittance, conductance and susceptance, respectively. $Z_0$ is the characteristic impedance of the resonator transmission line. Further, we neglect the actual tip resistance $R_t$ (Fig. 1(c)). Provided that $|z| \ll 1$, one gets:

$$\frac{\partial \mathrm{Arg}(\Gamma_t)}{\partial c_i} \approx -4\zeta\eta\frac{\partial \zeta}{\partial c_i} - 2\frac{\partial \eta}{\partial c_i}, \tag{6}$$

where $\zeta = \mathrm{Re}\, z$, and $\eta = \mathrm{Im}\, z$, and if:

$$\left|\zeta\eta\frac{\partial \zeta}{\partial c_i}\right| \ll \left|\frac{\partial \eta}{\partial c_i}\right|, \tag{7}$$

one can write for the frequency shift:

$$\Delta\omega_r \approx -2Z_0 f_0 \sum_i \frac{\partial B_t}{\partial c_i} dc_i. \tag{8}$$

And finally, expressing the tip susceptance through the effective tip capacitance $B_t = \omega C_t$, we obtain:

$$\frac{\Delta\omega_r}{\omega_r} \approx -2Z_0 f_0 \sum_i \frac{\partial C_t}{\partial c_i} dc_i. \tag{9}$$

Note that as follows from the last expression, the relative frequency shift does not depend on frequency directly in the approximations used here. Hence the frequency dependence of the relative frequency shift appears only through the frequency dependence of the properties of the tip-sample system.

Direct measurements and calculations show that the conditions of Eqs. (3) and (7) setting the validity range for the approximations are fulfilled in our microscope. Being equipped with Eq. (9), we can analyze the frequency response of the microscope using the equivalent circuit Fig. 1(c) with $\{c_i\}=\{C_{str}, C_{sc}, C_s, R_s\}$. The change of the effective tip capacitance due to a sample is:

$$dC_t = \sum_i \frac{\partial C_t}{\partial c_i} dc_i = \frac{\partial C_t}{\partial C_{str}} dC_{str} + C_p(f), \qquad (10)$$

where $C_p(f)$ is the frequency-dependent parallel capacitance of the simpler circuit consisting of only three elements $C_{sc}$, $C_s$, and $R_s$, as shown in the inset of the upper panel of Fig. 4. This circuit, with well-known properties, allows us to reproduce all essential features of the resonator frequency behavior in the region of validity of the electrostatic approximation. The change of the effective tip conductance $G_p(f)$ due to a sample can be introduced similar to $C_p(f)$. For the following analysis of the frequency response of the microscope, we neglect the stray capacitance $C_{str}$ because it is a frequency-independent additive to the parallel capacitance $C_p(f)$. The phase response of the circuit determined mainly by the value of the parallel capacitance $C_p(f)$, governs the phase change of the wave in the transmission line upon reflection from the tip, and hence, the resonant frequency shift. In the Appendix it is shown that the effective tip conductance $G_p(f)$ determines the change of the $Q$-factor of the resonant modes.

The frequency behavior of $C_p$ and $G_p/(2\pi f)$ with frequency-independent values of $C_{sc}$, $C_s$, and $R_s$ is shown in Fig. 4 in a dimensionless coordinate system. The capacitance and conductance are measured in values of $C_{sc}$, and the unit of frequency $f$ is $(2\pi)^{-1}$. As can be seen from Fig. 4, at frequencies much higher than the pole frequency defined by the expression $\omega(C_{sc} + C_s)R_s = 1$, the value of $C_p$ is very weakly dependent on $R_s$ and the conductance contrast becomes "invisible" for the microwave frequency shift. The transformation of the contrast of the images of Fig. 3 with increasing frequency can be explained by this fact. At high enough frequencies and under the condition $C_s \ll C_{sc}$, the microscope response becomes sensitive only to the dielectric properties of the sample. Optimal frequencies for imaging of particular properties can be found with use of Eqs. (9) and (A.5) as points of the maximum absolute values of partial derivatives over the equivalent circuit parameters.

*c) quantitative analysis*

Figure 5 shows relative frequency shift referenced to the bare LAO substrate versus frequency for four different locations around the milled trench. The notation for different parts of the film correspond to those of Fig. 3. Point "D" with a strong dependence of the frequency shift on frequency below about 6 GHz is close to the island corner and it received the largest Ga dose outside the trench. The dashed lines in Fig. 5 are guides to the eye. The factors with the largest contribution to the measurement error are the calibration error due to incomplete representation of the actual tip shape by numerical calculations made in the probe calibration process and the progressive damage of the tip during scanning. The overall measurement error due to these factors was estimated by numerical calculations and is reflected by the error bars in the plot.

The curved solid (green) line in Fig. 5 is the frequency shift relative to the bare LAO substrate calculated using Eq. (9) for the tip-sample system model of Fig. 1(c) with the following parameters: $Z_0$=50 Ω, $C_{sc}$=50 fF, $C_{str}$ =11.7 fF, $C_s$=5.7 fF, $R_s$=10 kΩ, $C_s^{LAO}$=3.9 fF, $R_s^{LAO}$ is equal to infinity. The last two values are for a bare LAO substrate. The curve corresponds to the most Ga-doped and, therefore, most conducting part of the film. The estimated pole frequency for this case is 285 MHz. The increase of the frequency shift observed at lower frequencies at the island corners corresponds in fact to the end of the slope of the $C_p$-vs.-frequency curve of Fig. 4. The horizontal straight solid (red) line in Fig. 5 is the result for the same model, but with $R_s$=10 MΩ, representing the undoped area. The value of $C_{str}$ was determined by measuring the resonant frequency shift after insertion of the probe tip into the open end of the resonator relative to the resonance frequency of the resonator without the tip. The value of $C_s^{LAO}$ was obtained by numerical finite element calculations, and the values of $C_s$ and $R_s$ were chosen to fit the experimental data, and they are on the order of magnitude expected.

As we see, Eq. (9) and the equivalent circuit Fig. 1(c) fairly well describe the measurement results up to a frequency of about 11 GHz. The experimental points at frequencies higher than 11 GHz deviate from the theoretical curve and show smaller relative frequency shifts. Two possible factors which might cause the observed behavior of the frequency shift at higher frequencies are the dispersion of the dielectric permittivity of the film, and the deformation of the probe tip during scanning.

Based on the measured frequency shifts, we have calculated the relative permittivity of the film and found that the relative permittivity of the intact BSTO film (see Fig. 5) is 550±70 at frequencies below 11 GHz. The relative permittivity of the damaged film (points "D" and "E", Fig. 5) is about 100 lower. The relative permittivity of the intact film at $f$ = 17.4 GHz would be close to 350 if it is calculated in the same way, which means an (unlikely) drop of 200 upon frequency change from 10.9 GHz to 17.4 GHz. This drop is too large and cannot be attributed to dispersion of the dielectric properties of the BSTO of any origin. Taking into account the data of Refs. [7] and [21], one should expect a change in the relative dielectric constant due to dispersion to be less than 100 in the whole frequency range from 1 GHz to 18 GHz. In fact, the dispersion of the BSTO properties could not be detected with our setup, and therefore, it is less than the measurement error of about 140, which is in agreement with Refs. [7] and [21]. The frequencies during the high-frequency experiment were changed in the following order: 17.4 GHz, 14.7 GHz, 8.7 GHz, 10.9 GHz, and 13.0 GHz. Hence, the downward trend seen in the curve at increasing frequency cannot be explained by a progressive change of the tip shape in the flow of the experiment, either.

We suggest that the deviation can be attributed to an enhanced contribution of radiated and evanescent excitations at the probe tip to the resonator response, which is not captured by the electrostatic numerical calculation. It should be noted that the relevant sample dimensions are not much smaller than the microwave wavelength $\lambda$ at the higher frequencies. Indeed, at a frequency of 15 GHz, for instance, $\lambda_{LAO} \approx 4$ mm in the LAO substrate with a relative permittivity $\varepsilon \approx 25$, whereas the substrate thickness $d \approx \lambda/8$ ($d$ = 0.5 mm). A noticeable deviation of the field structure in the film from the static one can be expected in this situation, and a full-wave analysis is needed to clarify the observed high-frequency behavior of the frequency shift.

Standard practical restrictions on the application of lumped-element models in the characterization of high frequency circuits is that the dimensions of the circuit elements should be less than $\lambda/20$. If this condition is applied to our microscope configuration, the lumped element model and the near-field electrostatic approximation can be used for measurements at frequencies below ~6 GHz only. Nevertheless, the broadband imaging still turns out to be useful in revealing the more conducting areas of the films as is evident from Fig. 3.

Finally, we would like to note that in future versions of the broadband microscope, frequency should be swept at each point (pixel) on the sample. The dielectric properties of the materials can be extracted then by fitting the frequency dependences for each sample point separately. One condition to realization of such an approach is design of a resonator-feed line coupling element capable of working in a wide frequency range with the coupling constant $\beta$ close to unity ($\beta = 1$ corresponds to the critical coupling; at $\beta = 1$ the reflection coefficient from the resonator at the resonance is equal to zero). Further, design of the tip-sample system in such a way that its pole frequency is close to the center of the frequency range covered by the microscope is another necessary condition for an effective realization of the broadband approach.

## ACKNOWLEGDMENTS

The work was supported by the University of Maryland/Rutgers NSF MRSEC under grant DMR-00-80008. We thank V. V. Talanov (Neocera, Inc.) and R. Ramesh for fruitful discussions, as well as C. Canedy and A. Stanishevsky for sample fabrication.

**References**

[*] Present address: Department of Chemistry, Duke University, Durham, NC, 27708.

[1] B. T. Rosner and D. W. van der Weide, Rev. Sci. Instrum. **73**, 2505 (2002).

[2] S. M. Anlage, V. V. Talanov, and A. R. Schwartz, in *Scanning Probe Microscopy: Electrical and Electromechanical Phenomena at the Nanoscale*, edited by S. V. Kalinin and A. Gruverman (Springer, New York, 2007).

[3] Y. Lu, T. Wei, F. Duewer, Y. Lu, N.-B. Ming, P. G. Schultz, and X. D. Xiang, Science **276**, 2004 (1997).

[4] K. S. Chang, M. A. Aronova, C. L. Lin, M. Murakami, M. H. Yu, J. Hattrick-Simpers, O. O. Famodu, S. Y. Lee, R. Ramesh, M. Wuttig, I. Takeuchi, C. Gao, and L. A. Bendersky, Appl. Phys. Lett. **84**, 3091 (2004).

[5] J. Lee, J. Park, A. Kim, K. Char, S. Park, N. Hur, and S. W. Cheong, Appl. Phys. Lett. **86**, 012502 (2005).

[6] V. V. Talanov, A. Scherz, R. L. Moreland, and A. R. Schwartz, Appl. Phys. Lett. **88**, 192906 (2006).

[7] K. S. Chang, M. Aronova, O. Famodu, I. Takeuchi, S. E. Lofland, J. Hattrick-Simpers, and H. Chang, Appl. Phys. Lett. **79**, 4411 (2001).

[8] V. Bobnar, P. Lunkenheimer, M. Paraskevopoulos, and A. Loidl, Phys. Rev. B **65**, 184403 (2002).

[9] D. C. Sinclair, T. B. Adams, F. D. Morrison, and A. R. West, Appl. Phys. Lett. **80**, 2153 (2002).

[10] V. V. Daniel, *Dielectric relaxation*. (Academic Press, London, New York, 1967).

[11] A. K. Jonscher, *Dielectric relaxation in solids*. (Chelsea Dielectrics Press, London, 1983).

[12] C. Gao, F. Duewer, and X. D. Xiang, Appl. Phys. Lett. **75**, 3005 (1999).

[13] C. Gao, F. Duewer, and X. D. Xiang, Appl. Phys. Lett. **76**, 656 (2000).

[14] C. Gao, B. Hu, P. Zhang, M. Huang, W. Liu, and I. Takeuchi, Appl. Phys. Lett. **84**, 4647 (2004).

[15] A. Imtiaz, M. Pollak, S. M. Anlage, J. D. Barry, and J. Melngailis, J. Appl. Phys. **97**, 044302 (2005).

[16] Z. Wang, M. A. Kelly, Z.-X. Shen, L. Shao, W.-K. Chu, and H. Edwards, Appl. Phys. Lett. **86**, 153118 (2005).

[17] D. E. Steinhauer, C. P. Vlahacos, F. C. Wellstood, S. M. Anlage, C. Canedy, R. Ramesh, A. Stanishevsky, and J. Melngailis, Appl. Phys. Lett. **75**, 3180 (1999).

[18] D. E. Steinhauer, C. P. Vlahacos, S. K. Dutta, F. C. Wellstood, and S. M. Anlage, Appl. Phys. Lett. **71**, 1736 (1997).

[19] D. E. Steinhauer, C. P. Vlahacos, F. C. Wellstood, S. M. Anlage, C. Canedy, R. Ramesh, A. Stanishevsky, and J. Melngailis, Rev. Sci. Instrum. **71**, 2751 (2000).

[20] D. M. Pozar, *Microwave Engineering*, 3rd ed. (J. Wiley, Hoboken, NJ, 2005).

[21] J. C. Booth, I. Takeuchi, and K.-S. Chang, Appl. Phys. Lett. **87**, 082908 (2005).

**APPENDIX**

The parallel tip capacitance $C_p$ determines the resonant frequency shift of the microscope. Here we discuss in more detail the role of the parallel tip conductance $G_p$ (inset Fig. 4) in the microscope response, in particular the quality factor $Q$. At a resonance, the ratio of the complex amplitudes of a partial wave within the resonator before, $A_1$, and after, $A_2$, one round trip between the coupling element and the tip can be expressed with use of a complex frequency $\hat{\omega} = \omega' + j\omega''$ as:

$$\frac{A_2}{A_1} = \exp j(\hat{\omega} T_0 + \varphi_t + \varphi_c) = \Gamma_t \Gamma_c \exp(-2\gamma L), \qquad (A.1)$$

where $T_0 = 1/f_0$ is the time of the round trip, and $\varphi_{t,c}$ are phase shifts of the wave upon reflection from the tip and the coupling element, respectively. It follows from the definition of $\gamma$ that $\omega' \equiv \omega$ and

$$\omega'' = f_0 \left( 2\alpha L - \ln|\Gamma_c| - \ln|\Gamma_t| \right) \qquad (A.2)$$

The $Q$-factor of a resonator can be written in the following form:

$$Q = \frac{\omega'}{2\omega''}. \qquad (A.3)$$

Similar to the steps leading to Eqs. (5) and (8), we obtain:

$$d\omega'' \approx 2Z_0 f_0 \sum_i \frac{\partial G_t}{\partial c_i} dc_i. \qquad (A.4)$$

For the validity of the last approximate expression, it is required that $\left| \zeta \eta \frac{\partial \eta}{\partial c_i} \right| \ll \left| \frac{\partial \zeta}{\partial c_i} \right|$, and it is assumed that $\omega'$ is fixed. $G_t \equiv G_p(f)$ of the circuit in Fig. 4, and for changes of losses due to the sample we have:

$$\Delta\left(\frac{1}{Q}\right) = 4 \frac{1}{\omega'} f_0 Z_0 \Delta G_p. \qquad (A.5)$$

The amplitude response of the circuit Fig. 1(c) is determined mainly by its parallel conductance $G_p$. The value $G_p/(2\pi f)$ directly determines the losses and the change of the resonator quality factor due to the sample. This value has a maximum at the pole frequency defined by the expression $\omega(C_{sc} + C_s)R_s = 1$, corresponding to the maximum of losses and to the minimum of the $Q$-factor.

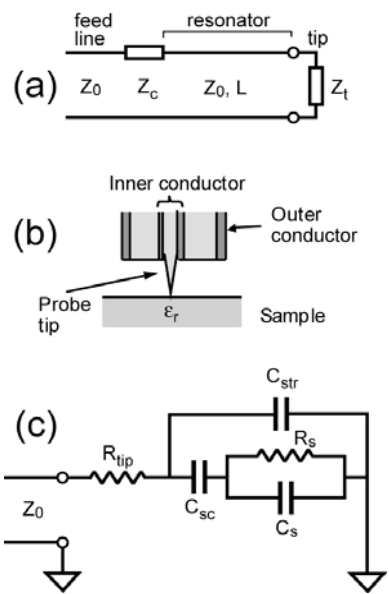

Fig. 1. (a) Electrical diagram of a transmission line resonator of length $L$ terminated by a tip with impedance $Z_t$; both the feed line and the resonator have a characteristic impedance $Z_0$, and the resonator is connected to the feed line through a coupling element with an impedance $Z_c$; (b) schematic of the open end of a coaxial resonator with a protruding probe tip and a sample; (c) simplified equivalent circuit describing the probe tip with a lossy dielectric sample nearby. $C_s$ and $R_s$ are the capacitance and resistance presented by the sample, which is coupled to the transmission line resonator through the resistance of the probe tip and the coupling capacitance $C_{sc}$. $C_{str}$ is the stray capacitance between the probe tip and the outer conductor.

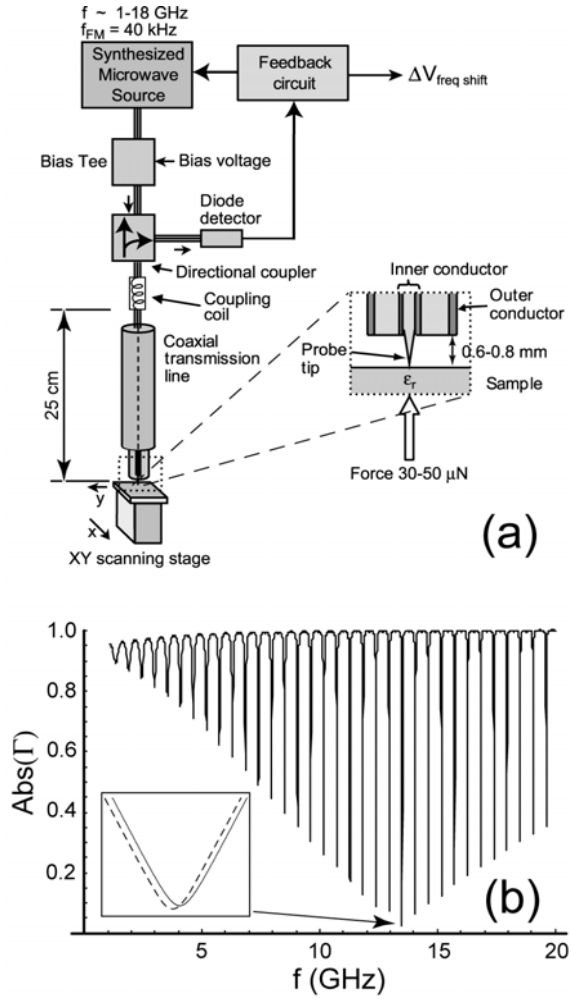

FIG. 2. (a) Schematic of the broadband near-field microwave microscope. The microwave signal is generated by the source, enters the resonator through the coupling coil, interacts with the sample, and is detected upon reflection outside the resonator. The source is frequency-modulated by the feedback circuit, and sent a dc correction signal proportional to the frequency shift of the microscope. Inset shows the tip-sample contact in detail. (b) Calculated absolute value of the reflection coefficient from the resonator, $\Gamma$, vs. frequency, $f$, for a purely inductive coupling between the feed line and the resonator. The length of the resonator used in the calculations was 25 cm, inductance of the coupling coil was 3 nH, and the characteristic impedance of both the feed line and the resonator was equal to 50 $\Omega$. Inset illustrates the scale of the resonance frequency shift due to a sample: solid line is the reflection coefficient of the resonant mode at the minimum of reflection without a sample, and the dashed line is the same with a sample at the probe tip.

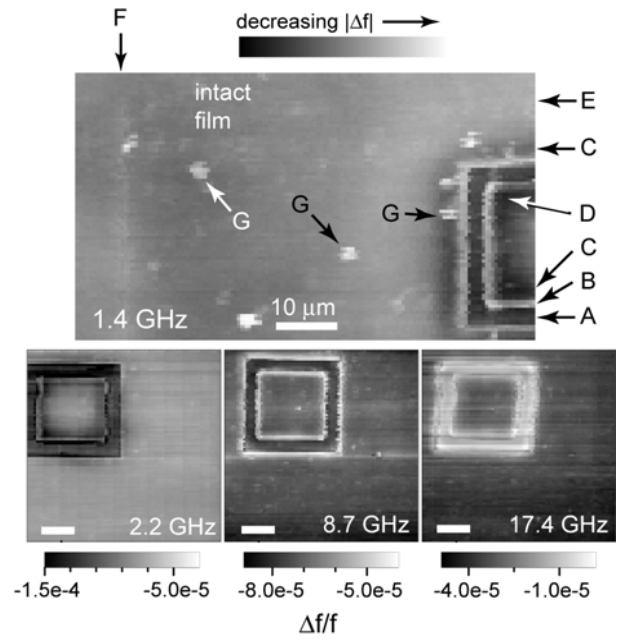

FIG. 3. Microwave frequency shift images of the FIB-milled BSTO film structure obtained at four different frequencies. The frequencies are shown in the corresponding images. Scale bars in the images of the lower panel correspond to 10 μm. The explanation of the features denoted with letters "A" through "G" is in the text.

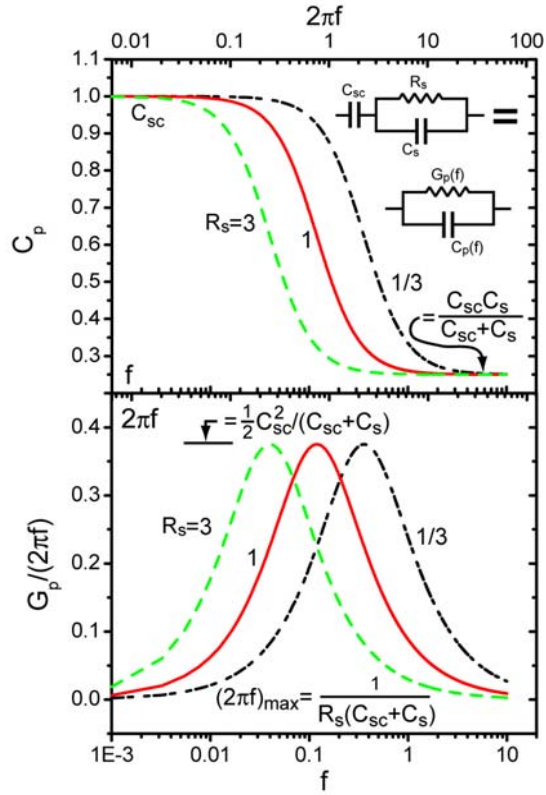

FIG. 4. (Color online) Calculated lumped element values $C_p$ and $G_p/(2\pi f)$ versus frequency curves for the circuit shown in the inset with frequency-independent $C_{sc}$, $C_s$, and $R_s$ values, shown in a dimensionless coordinate system. The capacitance and conductance are measured in units of $C_{sc}$, and the unit of frequency $f$ is $(2\pi)^{-1}$. Each panel contains three curves corresponding to three different sample resistances $R_s = 1/3$ (dashed, green), 1 (solid, red), 3 (dot-dashed, black).

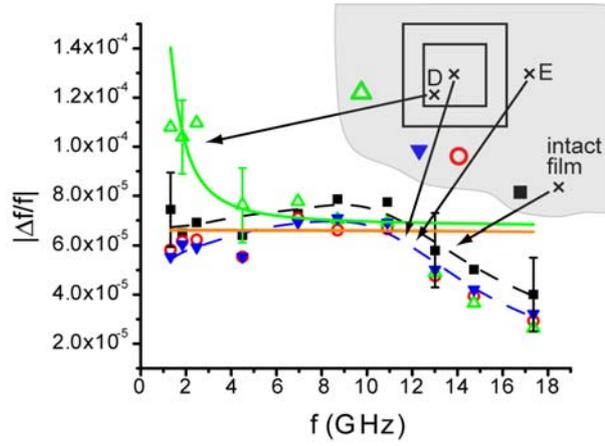

FIG. 5. (Color online) Relative frequency shift $\Delta f/f$ (referenced to the bare LAO substrate) vs. frequency for four different locations around the milled trench (refer to Fig. 3): black solid squares (■) correspond to an intact part of the film; (○) red hollow circles correspond to a damaged film with small Ga doping (denoted "E" in Fig. 3); (▼) blue solid triangles correspond to the middle of the island; and (△) green hollow triangles correspond to areas close to the island corners (denoted "D" in Fig. 3). The curved solid (green) line is the frequency shift relative to the bare LAO substrate calculated for the tip-sample system model of Fig. 1(c) with the following parameters: $Z_0=50$ Ω, $C_{sc}=50$ fF, $C_{str}=11.7$ fF, $C_s=5.7$ fF, $R_s=10$ kΩ, $C_s^{LAO}=3.9$ fF, $R_s^{LAO}$ is equal to infinity. The last two values are for a bare LAO substrate. The horizontal straight solid (red) line is the result of the same calculation with $R_s=10$ MΩ. The dashed (black and blue) lines are guides to the eye.